\documentclass[useAMS,usenatbib]{mn2e}

\usepackage{natbib}
\usepackage{graphicx}
\usepackage{amssymb}
\usepackage{color}
\usepackage{caption}

\title[UMBH feedback in compact galaxies]{Ultra-massive black hole feedback in compact galaxies}
\author[ ]
{W. Ishibashi$^{1}$\thanks{E-mail: wako.ishibashi@physik.uzh.ch} and A. C. Fabian$^{2}$
\footnotemark[0]\\
$^{1}$Physik-Institut, Universitat Zurich, Winterthurerstrasse 190, 8057 Zurich, Switzerland 
\footnotemark[0]\\
$^{2}$Institute of Astronomy, Madingley Road, Cambridge CB3 0HA 
}

\begin{document}

\pdfminorversion=4

\date{Accepted ? Received ?; in original form ? }

\pagerange{\pageref{firstpage}--\pageref{lastpage}} \pubyear{2012}

\maketitle

\label{firstpage}

\begin{abstract} 
Recent observations confirm the existence of ultra-massive black holes (UMBH) in the nuclei of compact galaxies, with physical properties similar to NGC 1277. The nature of these objects poses a new puzzle to the `black hole-host galaxy co-evolution' scenario. We discuss the potential link between UMBH and galaxy compactness, possibly connected via extreme active galactic nucleus (AGN) feedback at early times ($z > 2$). In our picture, AGN feedback is driven by radiation pressure on dust. We suggest that early UMBH feedback blows away all the gas beyond a $\sim$kpc or so, while triggering star formation at inner radii, eventually leaving a compact galaxy remnant. Such extreme UMBH feedback can also affect the surrounding environment on larger scales, e.g. the outflowing stars may form a diffuse stellar halo around the compact galaxy, or even escape into the intergalactic or intracluster medium. On the other hand, less massive black holes will drive less powerful feedback, such that the stars formed within the AGN feedback-driven outflow remain bound to the host galaxy, and contribute to its size growth over cosmic time. 
\end{abstract}



\section{Introduction}

Scaling relations between supermassive black holes and their host galaxies, such as the black hole mass-stellar velocity dispersion ($M_{BH} - \sigma$) or black hole mass-bulge mass ($M_{BH} - M_b$) relations, have been extensively studied over the past decades \citep[e.g.][and references therein]{Kormendy_Ho_2013}. These observational correlations point towards some form of `co-evolution', in which the central black hole is connected to its host galaxy, through the active galactic nucleus (AGN) feedback \citep{Silk_Rees_1998, Fabian_1999, King_2003}. 

The recent discovery of ultra-massive black holes (UMBH with $M_{BH} \gtrsim 10^{10} M_{\odot}$) in compact galaxies (with sizes $R \lesssim 2$kpc), poses a new puzzle to the co-evolutionary scenario. The first example is NGC 1277 in the Perseus cluster, which is reported to have a central black hole of mass $M_{BH} \sim 1.7 \times 10^{10} M_{\odot}$ located at the centre of a compact galaxy with a half-light radius of $R_e \sim 1.6$ kpc \citep{vandenBosch_et_2012}. The corresponding black hole-to-total stellar mass ratio is of the order of $\sim 14\%$, much larger than the usually expected values. We note that there are some uncertainties concerning the black hole mass estimate in NGC 1277 \citep{Graham_et_2016, Scharwachter_et_2016}; nonetheless, the system still harbours one of the most massive black holes known, remaining a major outlier in the black hole-host galaxy scaling relations \citep[][]{Walsh_et_2016}. Recent observations confirm the existence of further sources in the local Universe, with physical properties very similar to NGC 1277 \citep{Ferre-Mateu_et_2017}. Interestingly, several UMBH candidates are found to reside in nearby galaxy clusters, implying very large mean black hole mass densities \citep{Fabian_et_2013}.  

What is unique about these systems is that exceptionally massive black holes are hosted in compact galaxies, suggesting an intriguing connection between the presence of a UMBH at the centre and the host galaxy's compactness. Here we consider the possibility of a direct link between UMBH and compact galaxy formation, through extreme AGN feedback at early times. In fact, powerful AGN feedback may be expected at high redshifts ($z \gtrsim 2$), when the accreting UMBH were likely shining as bright quasars. 

It is now becoming clear that the effects of AGN feedback on its host galaxy can be quite complex, with both positive and negative feedback playing a role \citep{Ishibashi_Fabian_2012, Silk_2013, Zubovas_et_2013}. We have previously discussed how AGN feedback, driven by radiation pressure on dust, may trigger star formation in the host galaxy \citep{Ishibashi_Fabian_2012}; and how such star formation induced in the feedback-driven outflow may account for the size evolution of massive galaxies observed over cosmic time \citep{Ishibashi_et_2013}. 

The possibility of star formation occurring within galactic outflows has now been observationally confirmed, with the first direct detection obtained for a nearby source \citep{Maiolino_et_2017}. 
In the following, we discuss the case of UMBH feedback and its potential effects on the host galaxy, both in terms of star formation triggering and gas removal.  


\section{AGN radiative feedback and triggered star formation}

\subsection{Radiative feedback}

We assume AGN feedback driven by radiation pressure on dust. Radiative feedback sweeps up the surrounding material into an outflowing shell. The equation of motion of the shell is given by:
\begin{equation}
\frac{d}{dt} [M_{g}(r) v] = \frac{L}{c} (1 + \tau_{IR} - e^{-\tau_{UV}}) - \frac{G M(r) M_{g}(r)}{r^2} \, ,
\label{Eq_motion}
\end{equation} 
where $L$ is the central luminosity, $M(r)$ is the total mass distribution, and $M_{g}(r)$ is the gas mass, assuming a thin shell approximation \citep{Thompson_et_2015, Ishibashi_Fabian_2015}.  
The infrared (IR) and ultraviolet (UV) optical depths are given by: $\tau_{IR,UV}(r) = \frac{\kappa_\mathrm{IR,UV} M_{g}(r)}{4 \pi r^2}$, where $\kappa_{IR}$ and $\kappa_{UV}$ are the IR and UV opacities, respectively.

The gravitational potential is modelled by a Navarro-Frenk-White (NFW) profile  
\begin{equation}
\rho(r) = \frac{\rho_\mathrm{c} \delta_\mathrm{c}}{(r/r_\mathrm{s})(1+r/r_\mathrm{s})^2} \, , 
\end{equation} 
where $\rho_\mathrm{c} = \frac{3 H^2}{8 \pi G}$ is the critical density, $\delta_\mathrm{c}$ is a characteristic density, and $r_\mathrm{s}$ is a characteristic scale radius. 
The corresponding mass profile is given by 
\begin{equation}
M(r) = 4 \pi \delta_\mathrm{c} \rho_\mathrm{c} r_\mathrm{s}^3 \, \left[ \ln \left(1+\frac{r}{r_\mathrm{s}}\right) - \frac{r}{r+r_\mathrm{s}} \right] \, .
\end{equation} 

The ambient gas density distribution can be parametrized as a power law of radius $r$, with slope $\alpha$:
\begin{equation}
n(r) = n_0 \left( \frac{r}{R_0} \right)^{-\alpha} \, , 
\end{equation}
where $n_0$ is the density of the external medium, and $R_0$ is the initial radius. 
The case $\alpha = 2$ corresponds to the isothermal distribution, $n(r) \propto 1/r^2$. 
This may be a good approximation in the inner regions, but should break down on larger scales, where the gas density is likely to fall off more steeply with radius. 
We thus consider a slightly modified form:
\begin{equation}
n(r) \propto \frac{1}{r^{2} (r+r_a)^{\gamma}} \, ,
\end{equation}
where $r_a$ is a scale radius and $\gamma > 0$. 
Below, we assume the case of $\gamma = 1$ and $r_a \sim$1 kpc. 
The corresponding gas mass is given by:
\begin{equation}
M_{g}(r) = 4 \pi m_p \int n(r) r^2 dr 
= K \ln \frac{r+r_a}{r_a} \, , 
\end{equation}
where $K = 4 \pi m_p n_0 R_0^2 (R_0+r_a)$. 

From Eq. \ref{Eq_motion}, a critical luminosity can be defined by equating the outward force due to radiation pressure and the inward force due to gravity: $L_\mathrm{E}^{'} = \frac{Gc}{r^2} M(r) M_\mathrm{{g}}(r) (1 + \tau_\mathrm{{IR}} - e^{-\tau_\mathrm{{UV}}})^{-1}$. This can be considered as a generalised form of the Eddington luminosity. The corresponding effective Eddington ratio ($\Gamma = L/L_E'$) is given by:
\begin{equation}
\Gamma = \frac{L r^2}{c G M(r) M_{g}(r)} (1 + \tau_{IR} - e^{-\tau_{UV}}) \, ,
\label{Eq_Edd_ratio}
\end{equation} 
which basically corresponds to the ratio of the radiative force to the gravitational force.

As the ambient medium is swept up by the passage of the outflowing shell, the resulting compression of the gas may induce local density enhancements, which in turn may trigger star formation \citep{Ishibashi_Fabian_2012}. As previously discussed, we adopt a simple parametrisation for the star formation rate (SFR) in the outflow: 
\begin{equation}
\dot{M}_{\star} = \epsilon_{\star} \frac{M_g(r)}{t_{flow}(r)}  \, , 
\label{Eq_SFR}
\end{equation}
where $\epsilon_{\star}$ is the star formation efficiency, and $t_{flow} = r/v(r)$ is the local flow time.  
 

\subsection{UMBH feedback-induced formation of compact galaxies}
\label{subsec_UMBH}

\begin{figure}
\centering
\begin{center}
\includegraphics[angle=0,width=0.4\textwidth]{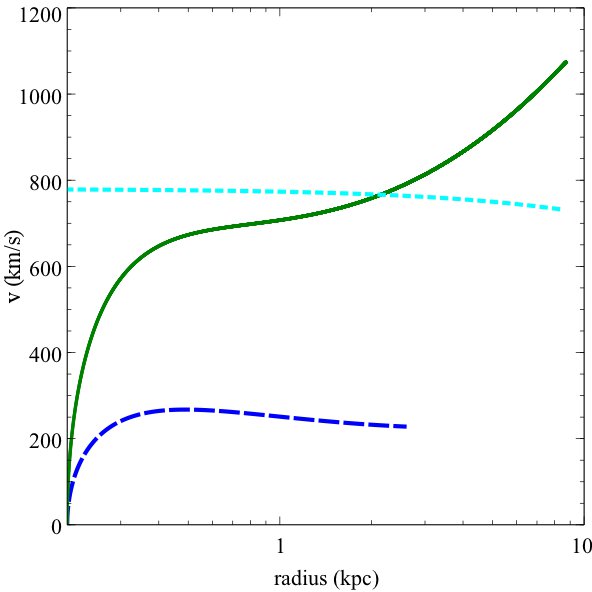} 
\caption{\small
Velocity as a function of radius: $L = 3 \times 10^{47}$erg/s, $n_0 = 5 \times 10^2 cm^{-3}$ (green solid); $L = 8 \times 10^{46}$erg/s, $n_0 = 2 \times 10^3 cm^{-3}$ (blue dashed); local escape velocity (cyan dotted). Fiducial parameters: $\kappa_{IR} = 3 cm^2/g$, $\kappa_{UV} = 10^3 cm^2/g$, $r_a = 1$kpc, $R_0 = 200$pc. 
}
\label{plot_v_r}
\end{center}
\end{figure}

Here we focus on the case of UMBH ($M_{BH} \sim 10^{10} M_{\odot}$, with standard bolometric Eddington luminosities of the order of $L_E \sim 10^{48}$erg/s), actively accreting at high redshifts. Figure \ref{plot_v_r} shows the radial velocity profile of the outflowing shell, derived by integrating the equation of motion (Eq. \ref{Eq_motion}). The shell velocity can be compared with the local escape velocity, which is defined as $v_\mathrm{esc} = \sqrt{2 \vert \Phi(r) \vert}$, where the gravitational potential is obtained by integrating the Poisson equation $\nabla^2 \Phi = 4 \pi G \rho$. We assume typical opacities of $\kappa_{IR} = 3 \mathrm{cm^2/g}$ and $\kappa_{UV} = 10^3 \mathrm{cm^2/g}$, with the initial radius set to $R_0 = 200$pc. We see that for a UMBH with a high central luminosity ($L = 3 \times 10^{47}$erg/s), the shell velocity exceeds the local escape velocity beyond $r \gtrsim 1$kpc (green solid curve); while for a less massive black hole, with a correspondingly lower central luminosity ($L = 8 \times 10^{46}$erg/s), the shell velocity is below the escape speed, and the outflowing shell remains bound to the galaxy (blue dashed curve). In Figure \ref{plot_v_r}, the escape velocity (cyan dotted line) is computed for NFW dark halo parameters appropriate for NGC 1277 \citep{Yildirim_et_2015}, while the escape speed may be somewhat different for the less massive black hole. But even adopting the conservative case of the isothermal potential, for which the escape velocity is given by $v_{esc} = 2 \sigma$, the corresponding speed is of the order of $v_{esc} \sim 500$km/s, which should not change the qualitative result.

\begin{figure}
\centering
\begin{center}
\includegraphics[angle=0,width=0.4\textwidth]{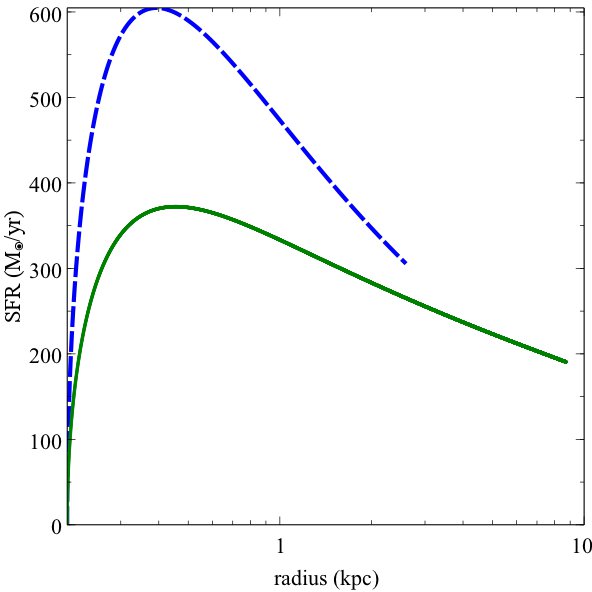} 
\caption{\small
Star formation rate as a function of radius, with $\epsilon_{\star} = 0.1$. Same physical parameters as in Figure \ref{plot_v_r}. 
}
\label{plot_SFR_r}
\end{center}
\end{figure}

As the radiation pressure-driven shell expands outwards, new stars can form within the outflowing shell. Figure \ref{plot_SFR_r} shows the associated star formation rates in the outflow, for a star formation efficiency of $\epsilon_{\star} = 0.1$. The newly formed stars are assumed to initially share the velocity of the outflowing shell in which they are formed (although, once decoupled from the shell, they will be slowed down by gravity). From Figure \ref{plot_v_r}, we observe that stars formed at small radii have low velocities, below the local escape speed, and thus remain bound to the galaxy; while stars formed at larger radii are born with higher velocities (since the shell is accelerated outwards). In the case of UMBH, the latter stars may exceed the escape speed on $\gtrsim$kpc scales, and hence be unbound. As a result, a compact galaxy remnant may be left behind. In contrast, for less massive black holes, with lower central luminosities, the newly formed stars never reach the escape speed, and thus remain bound to the galaxy.  

We note that the prescription for the star formation rate (Eq. \ref{Eq_SFR}) is simply a parametrisation of the overall rate of conversion of gas into stars within the outflowing shell, induced by AGN feedback; and Figure \ref{plot_SFR_r} shows the corresponding radial profile (SFR(r) at radius $r$). The total stellar mass added in the process can be obtained by integrating the star formation rate over the AGN feedback timescale, and assuming some value for the star formation efficiency. There are several observational indications that the star formation efficiency increases at high redshifts, and is higher in starburst-like systems \citep{Tacconi_et_2017, Scoville_et_2017}. 
Assuming a characteristic timescale ($\sim 10^7$yr, shorter than the Salpeter time), we obtain an order-of-magnitude estimate for the stellar mass, from a few times $\sim 10^9 M_{\odot}$ possibly up to $\sim 10^{10} M_{\odot}$. 
This is to be compared with the typical host stellar mass ($M_{\star} \sim 10^{11} M_{\odot}$) of compact galaxies \citep[e.g.][]{Ferre-Mateu_et_2017}. We see that the increase in stellar mass due to AGN feedback-triggered star formation is not dramatic, but may still form a non-negligible fraction of the total stellar mass. 
In cases where the gas reservoir is not completely removed by AGN feedback, several re-accretion events can occur at later times (with associated feedback-driven star formation episodes), further contributing to the growth of the host galaxy \citep{Ishibashi_et_2013}. 

The formation of a compact galaxy requires a very massive black hole, with a very high luminosity, capable of accelerating the shell to escape speed. In physical terms, an increase in the luminosity implies a higher effective Eddington ratio (Eq. \ref{Eq_Edd_ratio}), which leads to efficient acceleration and resulting high velocity. For a given central luminosity, a decrease in the ambient density will naturally facilitate the shell escape. Provided that the black hole is massive enough and that the surrounding gas density is not too high, UMBH feedback may blow away all the gas beyond a $\sim$kpc-scale, eventually leaving a compact galaxy remnant. On the other hand, if the black hole is less massive and the ambient gas density is high, the stars formed in the outflowing shell remain bound to the galaxy, contributing to the development of its spheroidal component and size evolution over cosmic time \citep{Ishibashi_Fabian_2014}. 
Furthermore, if the outflowing shell remains trapped in the galactic halo, gas may later fall back, providing fuel for further accretion and star formation.

Therefore, extreme UMBH feedback coupled with moderate ambient densities, may favour the formation of compact galaxies;  while less massive black holes in higher density environments may drive weaker feedback, resulting in more `normal' extended galaxies. 
The central luminosity may play a major role in determining the radial velocity profile of the outflowing shell, and hence the galaxy compactness; while, the ambient density may be the dominant factor in setting the triggered star formation rate 
(see Appendix \ref{Sect_Appendix}).


\section{Discussion}

Observations have uncovered a number of compact UMBH host galaxies in the local Universe, which share similarities with NGC 1277  \citep{vandenBosch_et_2012, Ferre-Mateu_et_2015, Ferre-Mateu_et_2017}. The UMBH host galaxies are characterised by uniformly old stellar populations (with ages $> 10$ Gyr), high metallicities, and high $[\alpha/Fe]$ abundance ratios \citep{Trujillo_et_2014, Ferre-Mateu_et_2017}. This indicates that the bulk of the stars formed in a short burst event at high redshift ($z > 2$), with correspondingly high star formation rates. The observed kinematics of compact galaxies is also peculiar, with centrally peaked velocity dispersions and high radial velocities ($v > 200$km/s) \citep{Martin-Navarro_et_2015, Ferre-Mateu_et_2017}. 
The set of particular features characterising the compact galaxies suggest that these are `relic galaxies', which remained basically unaltered since their formation at high redshift, i.e. without undergoing merger episodes at later times \citep{Trujillo_et_2014, Ferre-Mateu_et_2015}.

In our picture, powerful UMBH feedback can blow away the surrounding material, eventually leaving a compact galaxy remnant. Such extreme feedback may be expected at high redshifts ($z > 2$), close to the peak epoch of both AGN and star formation activities, when UMBH were rapidly accreting (e.g. on timescales comparable to the Salpeter time). Since the gas reservoir is removed by early UMBH feedback, the formation of new stars is inhibited at later times. However, star formation may be triggered within the AGN feedback-driven outflow itself. As the high-speed outflow sweeps through the galaxy on a short timescale, a strong radial stellar age gradient is not expected. This may be compatible with the uniformly old stellar populations observed in compact galaxies \citep{Trujillo_et_2014, Ferre-Mateu_et_2017}. Stars formed in the outflowing shell are also expected to have high radial velocities; radial velocities of several hundred km/s are observed in compact galaxies \citep{Martin-Navarro_et_2015, Ferre-Mateu_et_2017}. 

Some of the outflowing stars, with the highest radial velocities, will be ejected from the galaxy, and even escape into the intergalactic or intracluster medium. Such UMBH feedback-ejected stars may form a diffuse stellar halo around the compact galaxy, also contributing to the so-called intracluster light (ICL). In fact, observations indicate that a significant fraction of stars in galaxy clusters is not bound to individual galaxies, but rather reside in a common diffuse component \citep{Coccato_et_2011, Melnick_et_2012, Zemcov_et_2014}. We recall that several compact galaxies are found in galaxy clusters, like NGC 1277, which is located in the core of the Perseus cluster \citep{Fabian_et_2013}. Recent HST observations seem to suggest that a significant population of intracluster stars is also present in the Perseus cluster \citep{Harris_Mulholland_2017}. 

In addition to the different physical mechanisms proposed for the formation of intergalactic stars \citep[e.g.][and references therein]{Tutukov_Fedorova_2011}, AGN feedback-driven stellar outflows may also play a role. The UMBH host galaxies could then be surrounded by extended distributions of stars. An interesting observational implication would be the direct detection of diffuse stellar haloes around compact galaxies, which may be achievable with future instruments. In this framework, the observational appearance of the source (e.g. the galaxy compactness) would be mainly determined by internal processes associated with the central UMBH, rather than environmental effects such as tidal stripping.

Outflowing stars may also be produced in wind-shock outflows, and this scenario has been investigated in numerical simulations \citep{Nayakshin_Zubovas_2012, Zubovas_et_2013}. In this picture, rapid cooling of the shocked gas leads to shell fragmentation, eventually triggering the formation of high-velocity stars on radial trajectories. In contrast, in the radiative feedback scenario, we do not expect the development of strong shocks, except possibly an outer shock in the swept-up material (which subsequently cools down). In fact, radiative cooling of dense gas is very efficient and radiation pressure acts on the bulk of the mass (the dust being mainly located in the innermost regions). Furthermore, recent observations indicate that large amounts of dust can be formed in core-collapse supernovae, with a significant fraction of the dust grains surviving in the medium, due to their large sizes \citep{Owen_Barlow_2015, Wesson_et_2015}. In our picture, fresh dust can be produced, when massive stars formed in the outflowing shell explode as supernovae. The release of such additional dust, spread into the surrounding environment, may contribute to sustain the overall feedback process. 
An important requirement in our model is the presence of dust at early times, since the whole AGN feedback process relies on radiation pressure on dust. Indeed, the large dust masses observed in submillimetre galaxies and quasars at $z > 4$ \citep[][and references therein]{Michalowski_et_2010} require prompt dust production, possibly by early supernovae. Rapid dust formation is also needed in order to account for the detection of dust at the highest redshifts ($z \sim 7$), indicating early enrichment  \citep{Watson_et_2015, Laporte_et_2017}. 

Following their fast formation event at high redshifts, the compact galaxies may evolve differently depending on the subsequent merger history. According to the popular `two-phase' galaxy formation scenario \citep[e.g.][]{Hilz_et_2013, Dubois_et_2013}, the first phase of in-situ star formation is followed by a sequence of merger episodes, which account for the increase in size observed from $z \sim 2$ to the present. In this scenario, the relic galaxies are galaxies that did not experience the second phase (due to the stochastic nature of mergers) and survived unchanged to the present day \citep{Trujillo_et_2014, Ferre-Mateu_et_2015, Ferre-Mateu_et_2017}. 

While the second growth phase is usually attributed to minor mergers, the observed size evolution of massive galaxies may also be explained by several episodes of star formation triggered by AGN feedback \citep{Ishibashi_et_2013}. 
In this context, we have previously discussed how such AGN feedback-driven star formation mainly contributes to the development of the spheroidal component of galaxies (with the most massive black holes associated with elliptical galaxies, while the smaller ones are linked to the central bulges of disc galaxies). In particular, we have shown that the relation between characteristic radius and mass (set by the action of radiation pressure on dust), of the form $R \propto \sqrt{M}$, may naturally account for the `mass-radius' scaling relation observed in early-type galaxies \citep{Ishibashi_Fabian_2014}.

Finally, we recall that compact galaxies hosting UMBH are found to be major outliers in the local black hole mass-bulge mass ($M_{BH}-M_b$) relation \citep{Ferre-Mateu_et_2015, Ferre-Mateu_et_2017}. In the case of UMBH feedback, the gas reservoir is removed at early times, such that further star formation is inhibited, and only a compact stellar bulge is left behind. At high redshifts, most UMBH host galaxies would then be similar to NGC 1277, with a high $M_{BH}/M_b$ ratio. There are observational indications that the $M_{BH}/M_b$ ratios are larger at higher redshifts compared to local values, suggesting a higher normalisation in the $M_{BH}-M_b$ relation at earlier cosmic epochs \citep[][and references therein]{Kormendy_Ho_2013}. Intriguingly, the stellar mass density profile of NGC 1277 is found to be in remarkable agreement with that of massive compact galaxies at high redshifts \citep{Trujillo_et_2014}, further supporting the `relic' nature of compact galaxies. 
As discussed in \citet{Fabian_et_2013}, the subsequent evolution may depend on the location within the galaxy cluster environment: if the compact galaxy sits at the centre of the potential well, it can re-accrete gas and stars, and may develop into a brightest cluster galaxy (BCG); whereas, if the galaxy is rapidly orbiting in the core of the cluster, it is unable to accrete further material and may survive as a compact remnant, like NGC 1277.

\section*{Acknowledgements }

We acknowledge Roberto Maiolino for helpful comments on the manuscript. 
We thank the anonymous referee for a constructive report. 

  
\bibliographystyle{mn2e}
\bibliography{biblio.bib}


\appendix

\section{Luminosity and density dependence}
\label{Sect_Appendix}

As discussed in Sect. \ref{subsec_UMBH}, the shell radial profiles are mainly determined by two parameters: the central luminosity and the ambient density. 
For a given density distribution, an increase in the luminosity leads to a smaller crossing radius (where the shell velocity exceeds the local escape velocity). In fact, a higher luminosity implies a higher effective Eddington ratio (Eq. \ref{Eq_Edd_ratio}), leading to efficient shell acceleration and resulting high velocity. 
Likewise, for a given central luminosity, a decrease in the external density leads to a smaller crossing radius, since the effective Eddington ratio increases with decreasing $n_0$ in the single scattering regime ($\Gamma_{SS} \propto 1/n_0$, cf Eq. \ref{Eq_Edd_ratio}).
Concerning triggered star formation, we note that an increase in the ambient density leads to higher star formation rates (directly scaling as $\dot{M}_ {\star} \propto n_0$); while a higher luminosity also implies higher star formation rates (as $\dot{M}_{\star} \propto v$).

\begin{figure}
\centering
\begin{center}
\includegraphics[angle=0,width=0.4\textwidth]{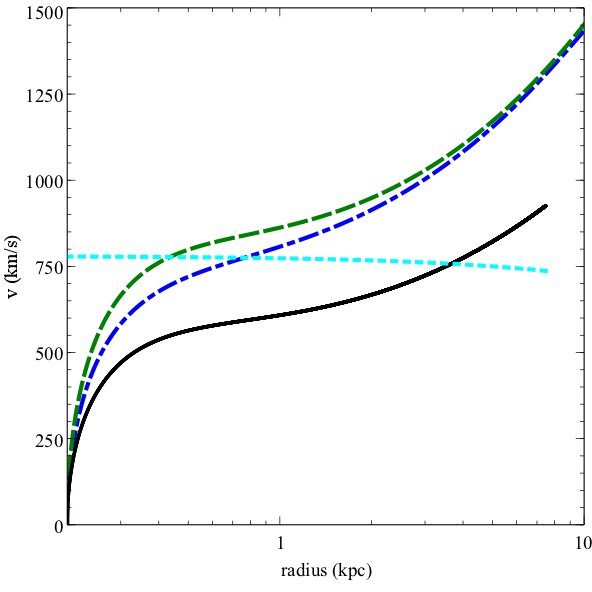} 
\caption{\small
Velocity as a function of radius. Variations in luminosity and density: $L = 5 \times 10^{47}$erg/s, $n_0 = 10^3 cm^{-3}$ (black solid, fiducial); $L = 10^{48}$erg/s, $n_0 = 10^3 cm^{-3}$ (green dashed); $L = 5 \times 10^{47}$erg/s, $n_0 = 5 \times 10^2 cm^{-3}$ (blue dash-dot); local escape velocity (cyan dotted).  
}
\label{plot_v_r_extra}
\end{center}
\end{figure}

\begin{figure}
\centering
\begin{center}
\includegraphics[angle=0,width=0.4\textwidth]{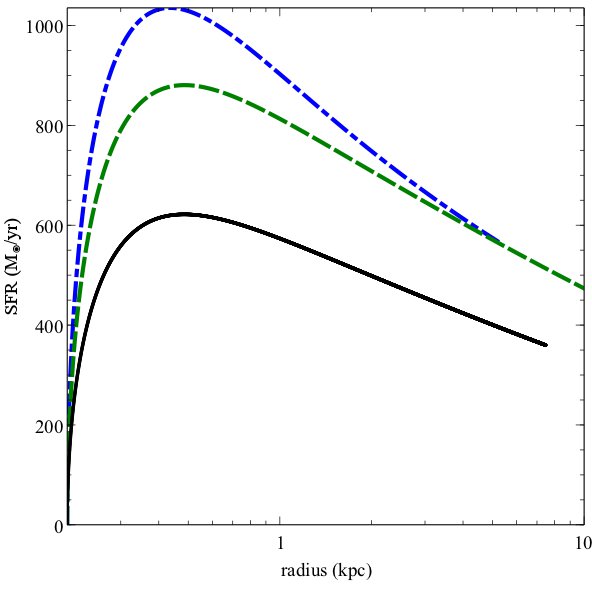} 
\caption{\small Star formation rate as a function of radius, with $\epsilon_{\star} = 0.1$. Same parameters as in Figure \ref{plot_v_r_extra}. 
}
\label{plot_SFR_r_extra}
\end{center}
\end{figure}

In Figures \ref{plot_v_r_extra} and \ref{plot_SFR_r_extra}, we quantify the dependence on the luminosity and ambient density, by varying both $L$ and $n_0$ by a factor of two.
From Figure \ref{plot_v_r_extra}, we see that a doubling in $L$ (green dash) has a larger effect on the velocity profile than a change by a factor of 2 in $n_0$ (blue dash-dot). Thus the central luminosity is more important in determining the crossing radius, and hence the galaxy compactness. Conversely, from  Figure \ref{plot_SFR_r_extra}, we observe that a doubling in $n_0$ (blue dash-dot) has a larger effect on the star formation rate than a doubling in $L$ (green dash). Thus the external density is the primary parameter governing the star formation rate. In fact, it is the combination of the two main parameters, luminosity and density, that ultimately determine the galaxy compactness and the importance of AGN feedback-triggered star formation.

\label{lastpage}

\end{document}